\documentclass[aps,prb,twocolumn,bibliography,amsmath,amssymb]{revtex4-2}
\usepackage{graphicx,xcolor}
\usepackage{bm}

\def\fig#1{Fig.~\ref{#1}}

\def\paragraph#1{}
\begin{document}
\title{Ferromagnetic Haldane state and dimer multiplet state of quantum ferromagnets}
\author{Shin \textsc{Miyahara}}
\email[]{smiyahara@fukuoka-u.ac.jp}
\affiliation{Department of Applied Physics, Fukuoka University, 8-19-1 Nanakuma, Jonan-ku, Fukuoka 814-0180, Japan}
\author{Isao \textsc{Maruyama}}
\affiliation{Department of Information and Systems Engineering, Fukuoka Institute of Technology, 3-30-1 Wajiro-higashi, Higashi-ku, Fukuoka 811-0295, Japan}

\begin{abstract}
  We present a theory of the realization of a ferromagnetic Haldane state in a spin-\(2\) bilinear-biquadratic spin system on an orthogonal-dimer chain. The coexistence of a ferromagnetic state and a Haldane state is due to the rigorous correspondence between the eigenstates of a spin-\(2\) model and a spin-\(\frac12\) Heisenberg model; i.e., ``eigensystem embedding.'' Numerical exact-diagonalization calculations indicate that the ground state in the model is a fractionally magnetized \(M=3/4\) Haldane state. Moreover, a ferromagnetic-dimer multiplet state is an exact ground state on a lattice, where the direct product of dimer singlet states is the ground state in a spin-\(\frac12\) Heisenberg model that includes one-, two-, and three-dimensional orthogonal-dimer lattices. Eigensystem embedding demonstrates that a quantum ferromagnet can be obtained for an arbitrary spin \(S \geqq 2 \) in any dimension and for any lattice in which anomalous ground states are realized in a spin-\(\frac12\) Heisenberg model.
\end{abstract}
\maketitle

\section{Introduction}

Quantum spin states have been attracting a wide range of researchers in physics and quantum computer science. To date, investigations of quantum spin states have been restricted to antiferromagnets; e.g., to entangled gapped quantum spin-liquid states~\cite{RevModPhys.89.040502}; Haldane gap states~\cite{PhysRevLett.50.1153}; Affleck-Kennedy-Lieb-Tasaki (AKLT) states~\cite{PhysRevLett.59.799}; symmetry-protected topological (SPT) states~\cite{RevModPhys.89.040502,PhysRevB.87.155114,arXiv2307.04788}; resonating-valence-bond (RVB) states~\cite{MaterResBull.8.153,Science.235.1196}; spin-liquid states~\cite{Nature.464.199,Science.332.1173,NatCommun.4.2287}; and direct products of dimer singlet states~\cite{JMathPhys.10.1063,PhysRevLett.47.964,PhysicaB+C.108.1069}. Recently, ``spin liquefaction'' of a ferromagnet has been proposed, in which an antiferromagnetic quantum state develops in a classical ferromagnetic background, in the spin-projection Hamiltonian, which includes a bilinear-biquadratic (BLBQ) model and this fractional ferromagnet can be regarded as the realization of entangled quantum states~\cite{arXiv2308.15372}.  The existence of such a quantum ferromagnet is ensured in any dimension and for any lattice because of the rigorous correspondence between a subset of the eigenstates in a spin-\(S\) model and the entire set of eigenstates in a spin-\(\frac12\) Heisenberg model. Since this correspondence can be regarded as ``eigensystem embedding,'' studies of quantum ferromagnets are attractive even in the context of quantum many-body scars~\cite{NatPhys.14.745,PhysRevB.98.235155,PhysRevB.98.235156,PhysRevLett.119.030601,PhysRevLett.124.180604}. In a quantum ferromagnet, a ferromagnetic and a quantum spin state coexist; consequently, a quantum spin state can be manipulated by using an external magnetic field.

\begin{figure}
\includegraphics[width=0.48\textwidth,bb= 0 0 608 351]{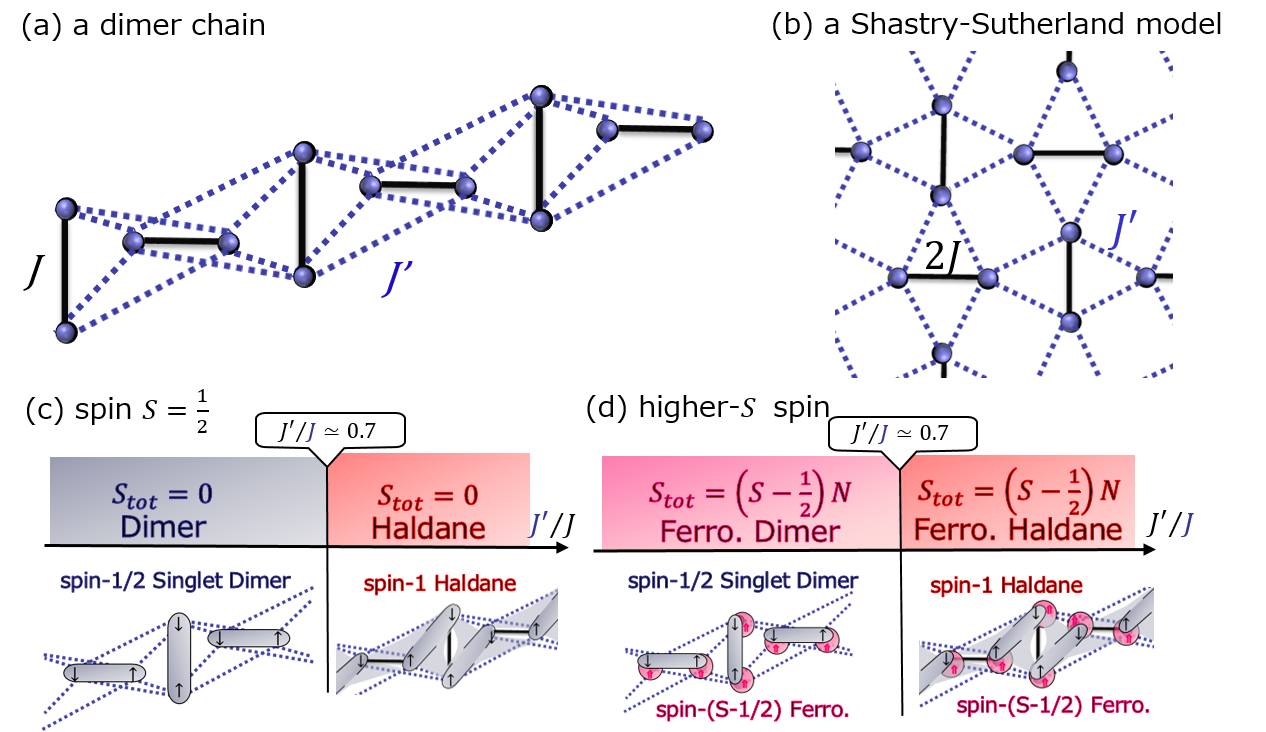}
\caption{(a) An orthogonal-dimer-chain model. (b) A two-dimensional orthogonal-dimer model, i.e., a two-dimensional Shastry-Sutherland Heisenberg model. (c) Phase transition from a direct product of dimer singlet states to a Haldane state in a spin-\(\frac12\) orthogonal-dimer-chain Heisenberg model. (d) Phase transition from a ferromagnetic-dimer septet state to a ferromagnetic Haldane state in a spin-\(2\) BLBQ orthogonal-dimer-chain model at \(\alpha_r\).}
\label{fig:model}
\end{figure}

A fractionally magnetized quantum spin state is rigorously an eigenstate at special points in a BLBQ model described by
\begin{align}
  H^{(S)} \left(\alpha\right)
  = & \sum K_{ij} \left[\cos \alpha \, {\bm S}_i \cdot {\bm S}_{j}
    + \sin \alpha \, \left({\bm S}_i \cdot {\bm S}_{j} \right)^2 \right].  &
  \label{eq:model-S} 
\end{align}
That is, rigorous eigenstate correspondence occurs at the points \(\alpha=\alpha_r, \alpha_r+\pi\) which are defined by
\begin{align}
  & \alpha_r = \pi-\arctan\left[ \frac{1}{2S(S-2)+1} \right],  &
  \label{eq:alpha-r}
\end{align}
for arbitrary \(S \geqq 1 \)~\cite{arXiv2308.15372}. Here, the eigenstates consisting of the spin states \(S\) and \(S-1\) in the BLBQ model are equivalent to the  eigenstates in a spin-\(\frac12\) Heisenberg model. The corresponding eigenstates can be obtained by using the interwiner \(\hat{C} = \prod_{i=1}^N \left( |S \rangle_i \langle \frac12 | + |S -1 \rangle_i \langle -\frac12 | \right) \), where \(\left| \frac12 \right\rangle \) and \(\left| -\frac12 \right\rangle \) are eigenstates of the \(z\) component of the spin operator \( s_i^z \) in the spin-\(\frac12\) model~\cite{arXiv2308.15372}. The corresponding eigenstate obtained by the interwiner \(\hat{C}\) from a spin-singlet state in the spin-\(\frac12\) model is a fractional ferromagnetic state, with \(S^z_{\rm tot} = \sum S_i^z = N \left( S - \frac12 \right), \) where \(N\) is the number of spins. The other degenerate states due to the ferromagnetism are obtained through the application of \( \left( S_{\rm tot}^- \right)^s \) [ \( s = 1, \dots, N(2S - 1) \)] to the corresponding states.

The rigorous eigenstate correspondence, i.e., eigensystem embedding, in the BLBQ model can be understood from the matrix elements of \(h_{ij}^{(S)} (\alpha) = K_{ij} \left[ \cos \alpha \, {\bm S}_i \cdot {\bm S}_j + \sin \alpha \, \left({\bm S}_i \cdot {\bm S}_j \right)^2 \right] \). The matrix elements \( \left\langle S \, S-2 \right| h_{ij}^{(S)} ( \alpha )  \left| S-1  \, S-1 \right\rangle \) and \( \left\langle S-2 \, S \right| h_{ij}^{(S)} ( \alpha )  \left| S-1  \, S-1 \right\rangle \) are given by
\begin{align}
  K_{ij} \left\{ \cos \alpha + \sin \alpha [ 2S(S-2)+1 ] \right\} \sqrt{2S^2-S},
\end{align}
which vanishes at \( \alpha = \alpha_r \) and \(\alpha_r + \pi \). At those points, the matrix of \(h_{ij}^{(S)} (\alpha_r) \) [\(h_{ij}^{(S)} (\alpha_r+\pi) \)] becomes a block-diagonal matrix in the subspace defined by \(S^z = S_i^z + S_j^z = 2S-2 \), and \( \left| S-1  \, S-1 \right\rangle\) becomes an eigenstate of \(h_{ij}^{(S)} (\alpha_r) \) [\(h_{ij}^{(S)} (\alpha_r+\pi) \)]. As a result, the blocks with the bases \( \left| S \, S \right\rangle, \left| S  \, S-1 \right\rangle, \left| S-1  \, S \right\rangle,\) and \( \left| S-1  \, S-1 \right\rangle\) at \( \alpha_r \) (\( \alpha_r + \pi \)) are given by
\begin{align}
   & (4S^2-2S)  \left(
   \begin{array}{cccc}
     \frac{J}{4}& 0 & 0 & 0 \\
     0 & -\frac{J}{4} &  \frac{J}{2} & 0 \\
     0 & \frac{J}{2} & -\frac{J}{4} & 0 \\
     0 & 0 & 0 & \frac{J}{4}
   \end{array} \right) - {\rm C} J \hat{I}_4, &
   \label{eq:h_ij_r}
 \end{align}
where \( J \equiv K_{ij} \sin \alpha_r \) [\( J \equiv K_{ij} \sin (\alpha_r + \pi) \)], \( C \equiv -\left( S^4 -4 S^3 +2S^2 -\frac{S}{2} \right) \), and \( \hat{I}_4 \) is the identity matrix. Equation~(\ref{eq:h_ij_r}) is equivalent to the matrix of a spin-\(\frac12 \) Heisenberg model~\cite{PhysRevB.12.3795}. As a result, the number of \( \left| S \right\rangle \)  and \( \left| S -1\right\rangle \) states is conserved for those states consisting only of \( \left| S \right\rangle \)  and \( \left| S -1\right\rangle \) in \(H^{(S)} (\alpha_r) \) [\(H^{(S)} (\alpha_r+\pi) \)] [Eq.~(\ref{eq:model-S}) at \( \alpha = \alpha_r (\alpha_r+\pi) \)], and a rigorous correspondence between the eigenstates of the spin-\( \frac12 \) Heisenberg model and those of \( H^{(S)} (\alpha_r) \)  [\( H^{(S)} (\alpha_r+\pi) \)] is realized. Moreover, when a corresponding state of a spin-singlet ground state in the spin-\( \frac12 \) Heisenberg model is a ground state of the BLBQ model, a fractionally magnetized quantum spin state can become stabilized in the BLBQ model; {\it i.e.}, a spin-singlet state in a spin-\( \frac12 \) model corresponds to a state with magnetization \( M  = 1 - \frac{1}{2S}  \) in \(H^{(S)} (\alpha_r) \), in which the ferromagnetic moment is reduced due to quantum fluctuations of the spins. In fact,  an \( M = 1 - \frac{1}{2S}  \) state is the ground state in a finite chain model at \(\alpha_r\) for  \( S \geqq 2 \)~\cite{arXiv2308.15372}.

In this paper, we study the ground state in a spin-\(2\) BLBQ model on a chain of orthogonal dimers~\cite{PhysRevB.52.12485,JPSJ.65.1387,EurPhysJB.15.227,JPSJ.87.123703} [\fig{fig:model}(a)] and two-dimensional orthogonal-dimer lattice, i.e., the Shastry-Sutherland lattice~\cite{PhysicaB+C.108.1069} [\fig{fig:model}(b)]. We discover anomalous fractionally magnetized states, i.e., a ferromagnetic Haldane state and a ferromagnetic dimer septet state [\fig{fig:model}(d)], where spin gap states realized in spin-\( \frac12 \) Heisenberg models [\fig{fig:model}(c)] and ferromagnetic state with the reduced magnetic moment coexist. The total spin moments \(S_{\rm tot} \) in the ferromagnetic Haldane state and the ferromagnetic dimer septet state are reduced from fully polarized state \(S_{\rm tot} = 2N \) to  \(S_{\rm tot} = \frac32 N\), where \(N\) is the total number of spins. Reflecting the coexistence of gapped and gapless states, the lowest excitation energy for the \(\Delta S_{\rm tot} = 1 \) process is finite, although the lowest excitation energy for the \(\Delta S_{\rm tot} = -1 \) process is gapless. These anomalous states are stabilized due to applications of the spin liquefaction of a ferromagnet~\cite{arXiv2308.15372} to frustrated-spin models.  

\section{quantum ferromagnet in spin-2 BLBQ models}

In the following, we restrict ourselves to a spin-2 BLBQ model. However, similar arguments are applicable to arbitrary spin-\( S\ (\geqq 2) \) models. The spin-2 BLBQ model at \(\alpha_r = \frac34 \pi \), with \(J_{ij} \equiv K_{ij} \sin \alpha_r \), is described by 
\begin{align}
  H^{(2)} ( \alpha_r )  = &  \sum J_{ij} \left[- {\bm S}_i \cdot {\bm S}_j
    + \left({\bm S}_i \cdot {\bm S}_j \right)^2 \right].  &
  \label{eq:model} 
\end{align}
Rigorous eigenstate correspondence occurs between a subset of the eigenstates in the spin-2 BLBQ model and the entire set of eigenstates in a spin-\(\frac12\) antiferromagnetic Heisenberg model in any dimension and for any lattice. When we consider a spin-\(\frac12\) Heisenberg model for which the ground state is anomalous, e.g., an SPT~\cite{RevModPhys.89.040502,PhysRevB.87.155114,arXiv2307.04788} or a spin-liquid~\cite{Nature.464.199,Science.332.1173,NatCommun.4.2287} system, the corresponding eigenstate of the BLBQ model is a ferromagnetic quantum state that has both an anomalous spin structure stabilized by quantum fluctuations and a ferromagnetic moment. However, whether the corresponding state is the ground state depends on the model.

\subsection{\(S = 2\) orthogonal-dimer chain}

Let us start from an \(S = 2\) orthogonal-dimer chain [\fig{fig:model} (a)], i.e., a highly frustrated ladder~\cite{PhysRevB.52.12485,JPSJ.65.1387,EurPhysJB.15.227,JPSJ.87.123703}, defined by
\begin{align}
  H^{(2)}_{\rm 1D} ( \alpha_r )  = &
  \sum_{x=1}^L J \left[- {\bm S}_{x, 1} \cdot {\bm S}_{x, 2}
    + \left( {\bm S}_{x, 1} \cdot {\bm S}_{x, 2} \right)^2 \right] & \nonumber \\
  & + \sum_{x=1}^L \sum_{i,j=1}^2 J^\prime \left[
    -{\bm S}_{x, i} \cdot {\bm S}_{x+1, j}  +  \left( {\bm S}_{x, i} \cdot {\bm S}_{x+1, j} \right)^2 \right],  &
  \label{eq:orthogonal-dimer} 
\end{align}
where \(J_{ij}\) on a dimer bond is defined as \(J\) and \(J_{ij}\) on inter-dimer bonds is defined as \(J^\prime\). We assume the periodic boundary conditions \( {\bm S}_{L+1,i} = {\bm S}_{1,i} (i = 1,2) \) unless specified otherwise. Here, \(L\) is the length of the chain, and the total number of spins is \(N=2L\). The corresponding states obtained from the entire set of eigenstates, which includes a Haldane state, in the spin-\(\frac12\) Heisenberg model \(h_{\rm 1D}\) are also eigenstates of the spin-2 model (\ref{eq:orthogonal-dimer}).

We summarize the features of the spin-\(\frac12\) orthogonal-dimer-chain model. The model is defined as
\begin{align}
  h_{\rm 1D} = &
  \sum_{x=1}^L J {\bm s}_{x, 1} \cdot {\bm s}_{x, 2}  + \sum_{x=1}^L \sum_{i,j=1}^2 J^\prime {\bm s}_{x, i} \cdot {\bm s}_{x+1, j}. &
  \label{eq:orthogonal-dimer-half} 
\end{align}
Here, \(J\) and \(J^\prime\) are exchange interactions and  \({\bm s}_{x,i}\)  is a spin-\(\frac12\) operator. The ground states of the spin-\(\frac12\) model (\ref{eq:orthogonal-dimer-half}) are a Haldane state (\(J^\prime/J \gtrsim 0.714\)) and a dimer singlet state (\(J^\prime/J \lesssim 0.714\))~\cite{EurPhysJB.15.227} [\fig{fig:model} (c)]. Note that the spin-\(\frac12\) model (\ref{eq:orthogonal-dimer-half}) is equivalent to a one-dimensional, spin-1 chain Heisenberg model,
\begin{align}
  H({\bm t}_x) = \frac{J}{2} \sum_{x=1}^L \left( {\bm t}_x^2 - \frac34 \right)
  + J^\prime \sum_{x=1}^L {\bm t}_x \cdot {\bm t}_{x+1},
\end{align}
where the operator \({\bm t}_x = {\bm s}_{x,1} + {\bm s}_{x,2} \) is defined at each dimer, and the Haldane state, which is the state \( t_x = 1 \), is identical to the ground state in a spin-1 Heisenberg chain~\cite{EurPhysJB.15.227}.

\begin{figure}
\includegraphics[width=0.48\textwidth,bb= 0 0 3704 3559]{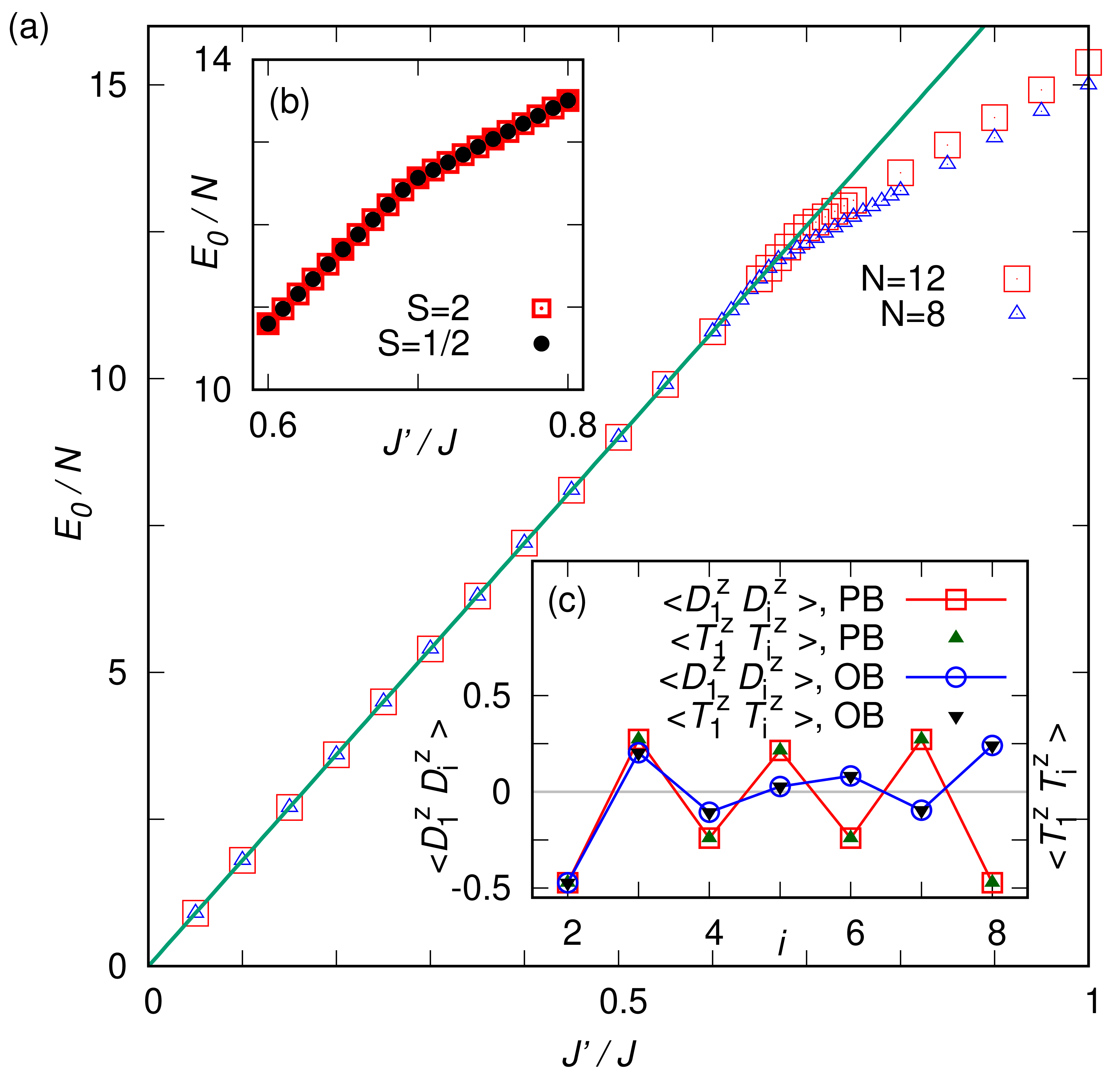}
\caption{(a) Ground-state energy per site, \(E_0/N\), as a function of \(J^\prime/J\) on 8- and 12-site clusters with \(S^z_{\rm tot} = 0 \) in a spin-\(2\) BLBQ orthogonal-dimer chain at \(\alpha_r\). (b) A comparison of \(E_0/N\) for a spin-\(2\) BLBQ model and for a spin-\(\frac12\) Heisenberg model on a 12-site cluster. (c) Dimer--dimer correlations \( \langle D_1^z D_i^z \rangle \) of the polarized state for \(J^\prime/J = 1\) in a spin-\(2\) BLBQ orthogonal-dimer chain on a 16-site cluster with periodic boundary (PB) conditions and open boundary (OB) conditions. This is equivalent to the spin--spin correlations \( \langle T_1^z T_i^z \rangle \) in a spin-\(1\) Heisenberg chain. }
\label{fig:E0_Jp}
\end{figure}

\begin{figure}
\includegraphics[width=0.48\textwidth,bb= 0 0 4042 3898]{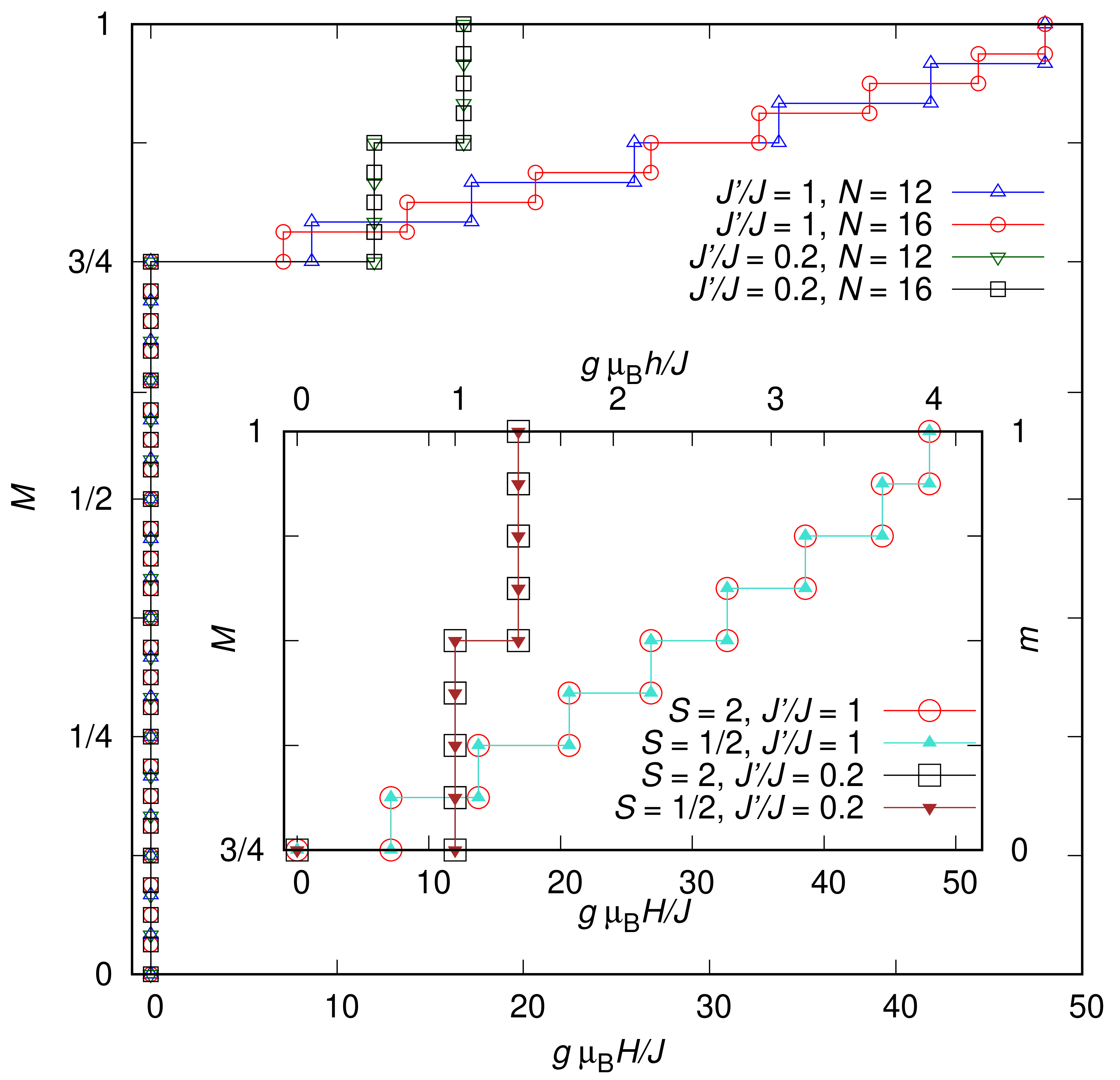}
\caption{Magnetization \(M \equiv S_{\rm tot}^z/2N \) for \(J^\prime/J = 0.2\) and 1 on 12- and 16-site clusters as a function of the external magnetic field \(g \mu_{\rm B} H/J \) in a spin-\(2\) BLBQ orthogonal dimer chain. (Inset) A comparison of the magnetization of the spin-\(2\) BLBQ model and the spin-\(\frac12\) Heisenberg model on a 16-site cluster. The magnetization of the spin-\(\frac12\) Heisenberg model, \(m \equiv s_{\rm tot}^z/(N/2) \), is plotted as a function of the external magnetic field \(g \mu_{\rm B} h/J \). }
\label{fig:M_H}
\end{figure}

In this way at least, a ferromagnetic Haldane state and a ferromagnetic dimer septet state are eigenstates of Eq.~(\ref{eq:orthogonal-dimer}). Note that in the dimer septet state, the total spin quantum number at each dimer is \(3\), and a polarized dimer state is \(\frac{1}{\sqrt{2}} \left( |2 1 \rangle - |1 2 \rangle \right) \). Numerical evidence is required to show that such a ferromagnetic quantum state can be the ground state. We therefore performed an exact diagonalization using the Lanczos method. The ground-state energy per site, \(E_0/N\), is shown in \fig{fig:E0_Jp}(a) as a function of \(J^\prime/J\) (\(0 < J^\prime/J \leqq 1 \)) on 8- and 12-site clusters in sectors  with \(S^z_{\rm tot} = \sum S_i^z = 0 \). We also calculated the total spin \(S_{\rm tot}\) of the ground state, and the result indicates that the ground state is a unique ferromagnetic state with the total spin  \(S_{\rm tot} = \frac32 N \) and that it has (\(3N+1 \))-fold degeneracy due to the ferromagnetic moment. Note that the lowest energies obtained using the Lanczos method are identical in sectors with \(S^z_{\rm tot} = 0, 1, \dots, \frac32 N \). As shown in Fig.~\ref{fig:E0_Jp}(b), the eigenenergies of the ground states are reproduced well by those of the spin-\(\frac12\) model described by Eq.~(\ref{eq:h_ij_r}). In these ways, fractionally magnetized states, i.e.,  a ferromagnetic Haldane state and a ferromagnetic dimer septet state, are ground states, and the quantum phase transition point is also identical to that in a spin-\( \frac12 \) system with \( \left(J^\prime/J \right)_c \sim 0.714 \) in a finite cluster [\fig{fig:model}(d)]. To clarify the identity of the polarized ferromagnetic Haldane states, we calculated the dimer--dimer correlation \( \langle D_1^z D_i^z \rangle = \langle (S_{1,1}^z + S_{1,2}^z - 3)(S_{i,1}^z + S_{i,2}^z - 3) \rangle \) and compared the value of \( \langle D_1^z D_i^z \rangle \) for an \(N\)-site cluster with the spin--spin correlation \( \langle T_1^z T_i^z \rangle \) of an \(N/2\)-site spin-1 Heisenberg chain. The results for \(J^\prime/J = 1\) and \(N = 16\) are shown in Fig.~\ref{fig:E0_Jp}(c). This figure shows that \( \langle D_1^z D_i^z \rangle \) and  \( \langle T_1^z T_i^z \rangle \) are identical for both periodic boundary conditions and open boundary conditions. We used the results in the sector with \(S_{\rm tot}^z = 18\) (\(S_{\rm tot}^z = 0\)) for the periodic boundary conditions and in the sector with \(S_{\rm tot}^z = 19\) (\(S_{\rm tot}^z = 1\)) for the open boundary conditions in the BLBQ (Heisenberg) model. We thus conclude that the polarized state of a ferromagnetic Haldane state is identical to a Haldane state that includes topological edge states. On the other hand, the polarized dimer septet states with \(S^z_{\rm tot} = \frac32 N \) are explicitly represented by
\begin{align}
  \left| \Psi_{\rm D} \right\rangle = & \prod_{\rm n.n.} \frac{1}{\sqrt{2}}
  \left( \left| 2 1 \right\rangle - \left| 1 2 \right\rangle \right)_{x},
  \label{eq:septet_dimer}
\end{align}
for which the eigenenergy is given by \( E_{\rm D} = 18  J^\prime N \). The entangled dimer septet state polarized with \( S_{\rm tot}^z= \frac32 N \) is explicitly represented by the \(S_z = 2\) and 1 spin states \(\left| 2 \right\rangle\) and \(\left| 1 \right\rangle\). In addition, the ground state in a finite-size cluster in an external magnetic field \(g \mu_{\rm B} H\) is also a corresponding state. The magnetization \(M \equiv S_{\rm tot}^z/2N \) in 12- and 16-site clusters for \(J^\prime/J = 0.2\) and 1 is shown in Fig.~\ref{fig:M_H}. The inset in Fig.\ref{fig:M_H} shows that the magnetization in the BLBQ model is identical to the magnetization \(m \equiv s_{\rm tot}^z/(N/2) \) in a spin-\(\frac12 \) model in an external magnetic field \(g \mu_{\rm B} h\), except for the shift due to the ferromagnetic component of the ferromagnetic quantum state and the modification of the magnitude of \(g \mu_{\rm B} h/J\) obtained from Eq.~(\ref{eq:h_ij_r}).  In the dimer septet phase, a magnetization plateau appears at the magnetization \( M = \frac78 \), as observed in the  spin-\(\frac12 \) model~\cite{EurPhysJB.15.227}. 

\begin{figure}
  \includegraphics[width=0.48\textwidth,bb= 0 0 480 480]{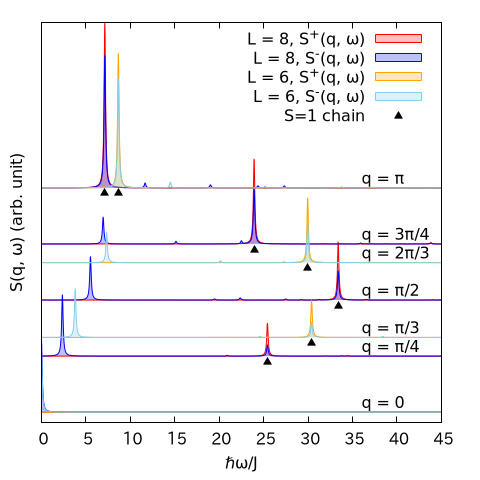}
  \caption{The dynamical structure factor \(S^{\pm} (q, \omega)\) for \(J/J^\prime = 1\) on 12- and 16-site clusters in a spin-\(2\) BLBQ orthogonal-dimer chain. The peak positions obtained from a spin-\(1\) Heisenberg chain are shown by triangles.} 
  \label{fig:Sq_H}
\end{figure}

Because both the Haldane state and the product of the dimer singlet states are spin-gapped states, the ferromagnetic Haldane state and the product of dimer septet states have spin-gap excitations, i.e., excitations for the process \(\Delta S_{\rm tot} = 1 \). The spin-gap energy can be read off from Fig.~\ref{fig:M_H}. The magnitude of the spin gap in a finite cluster is identical to that in the spin-\(\frac12\) model. The magnetic excitation in the ferromagnetic Haldane phase can be obtained from the dynamical structure factor,
\begin{align}
  S^{\pm} (q, \omega) = -\frac{1}{\pi}
  {\rm Im} \sum_n \frac{\left| \langle n | S^\pm (q) | 0 \right\rangle|^2}{\hbar \omega - \hbar \omega_{n0} + i \epsilon},
\end{align}
where \( S^\pm (q) = \sum \left( S^\pm_{x,1}  + S^\pm_{x,2} \right) e^{i q x} \). Here, \(|0\rangle\) is the ground state, which has the eigenenergy \(E_0\), and \(|n\rangle\) is the \(n\)-th magnetic excitation state, with the eigenenergy \(E_n\). In addition, \( \hbar \omega_{n0} = E_n - E_0 \) and \( \epsilon/J = 0.1 \). We calculated \( S^{\pm} (q, \omega)\) for \(J^\prime/J = 1\) on 12- and 16-site clusters via a continued-fraction expansion using the Lanczos method~\cite{RevModPhys.66.763}. We fixed the number of Lanczos steps to be 1000, which achieved convergence of  \( S^{\pm} (q, \omega)\) in the energy range \( 0 \leqq \hbar \omega/J \leqq 45 \). The results are shown in Fig.~\ref{fig:Sq_H}. The lowest energy \( \Delta S_{\rm tot} = 1 \) branch of \( S^+ (q) \) in the ferromagnetic Haldane state is identical to the lowest triplet branch in the Haldane state of an \(S=1\) chain model. In addition, a spin-wave-like branch (\( \Delta S_{\rm tot} = -1 \)) exists, which reflects the ferromagnetic component. Note that we calculated the total spin of each excited state using the conventional Lanczos method.  The magnitude of the spin-gap energy \( \Delta E_{\rm H} \) in the ferromagnetic Haldane phase can be estimated in the thermodynamic limit to be \( \Delta E_{\rm H} \sim 12 \times  0.41 J^\prime \) from that of an \(S=1\) spin chain~\cite{PhysRevB.33.659}. The spin-gap energy in the dimer septet phase is exactly given by \( \Delta E_{\rm D} =  12 J \), and nonet excitations with  \( S_{\rm tot} = \frac32 N + 1 \) and \( S_{\rm tot}^z = \frac32 N + 1 \) are given by \( \left| \Psi_{\rm D}^+ \right \rangle = \left| 2 2 \right \rangle_{y}  \prod_{x \neq y} \frac{1}{\sqrt{2}} \left( \left| 2 1 \right\rangle - \left| 1 2 \right\rangle \right)_{x}. \) The nonet excitations have a localized character, as does the triplet excitations in the spin-\(\frac12 \) model, which leads to the flat-band excitation \( \Delta S_{\rm tot} = 1 \) and  a magnetization plateau. As in the ferromagnetic Haldane phase, a spin-wave-like branch (\( \Delta S_{\rm tot} = -1 \)) exists that reflects the ferromagnetic component.

\subsection{\(S = 2\) Shastry-Sutherland lattice}

A quantum ferromagnet can be realized even in higher-dimensional lattices. As a typical example of two-dimensional models, let us consider a two-dimensional Shastry-Sutherland lattice [Fig.~\ref{fig:model}(b)],
\begin{align}
  H^{(2)}_{\rm 2D} ( \alpha_r )  = &
  \sum_{n.n.} 2J \left[- {\bm S}_{i} \cdot {\bm S}_{j}
    + \left( {\bm S}_{i} \cdot {\bm S}_{j} \right)^2 \right] & \nonumber \\
  & + \sum_{n.n.n.} J^\prime \left[
    -{\bm S}_{i} \cdot {\bm S}_{j}  +  \left( {\bm S}_{i} \cdot {\bm S}_{j} \right)^2 \right],  &
  \label{eq:2d-orthogonal-dimer} 
\end{align}
where interactions on a dimer bond are defined as \(2J\) and interactions between dimers are defined as \(J^\prime\). Even in a two-dimensional model, the corresponding states obtained from the entire set of eigenstates in the spin-\(\frac12\) Heisenberg model \(h_{\rm 2D}\) are also eigenstates of the spin-2 model (\ref{eq:2d-orthogonal-dimer}).

A spin-\( \frac12 \) Heisenberg model on a Shastry-Sutherland lattice~\cite{PhysicaB+C.108.1069} is described as
\begin{align}
  h_{\rm 2D} = & \sum_{n.n.} 2J {\bm S}_{i} \cdot {\bm S}_{j}  + \sum_{n.n.n.} J^\prime {\bm S}_{i} \cdot {\bm S}_{j}, 
  \label{eq:2d-orthogonal-dimer-half} 
\end{align}
where interactions on a dimer bond are defined as \(2J\) and interactions between dimers are defined as \(J^\prime\). It is well known that a direct product of dimer singlet states on \(2J\) bonds is a rigorous eigenstate and a ground state when \(J^\prime/2J < 0.765\)~\cite{PhysRevB.87.115144}. As shown in Ref.~\cite{PhysicaB+C.108.1069}, the Shastry-Sutherland lattice can be represented by a sum of isosceles triangles.

The spin-\( 2 \) BLBQ model (\ref{eq:2d-orthogonal-dimer}) is also regarded as the sum of the spin-\( 2 \) BLBQ model on an isosceles triangle:
\begin{align}
  H_{\rm iso}^{(2)} (\alpha_r) = & J \left[ -  {\bm S}_{i} \cdot {\bm S}_{j} + \left({\bm S}_{i} \cdot {\bm S}_{j}\right)^2  \right] & \nonumber \\
  & + J^\prime \left[ - {\bm S}_{i} \cdot {\bm S}_{k} + \left({\bm S}_{i} \cdot {\bm S}_{k}\right)^2 \right] & \nonumber \\
  & + J^\prime \left[ - {\bm S}_{j} \cdot {\bm S}_{k} + \left({\bm S}_{j} \cdot {\bm S}_{k}\right)^2 \right]. &
\end{align}
Since the dimer bonds are shared between two isosceles triangles, we define the interaction on a dimer bond as \(2J\). When \( 0 < J^\prime/J < 1 \), two of the 20 ground states of \( h_{\rm iso}^{(2)} (\alpha_r) \) (\(S_{\rm tot} = 4\) and \(5\) states) are given by
\begin{align}
  \frac{1}{\sqrt{2}} \left( \left| 2 \,\,\, 1  \right\rangle_{ij} - \left| 1 \,\,\, 2  \right\rangle_{ij} \right) \left| 2 \right\rangle_k, \\
  \frac{1}{\sqrt{2}} \left( \left| 2 \,\,\, 1 \right\rangle_{ij} - \left| 1 \,\,\, 2 \right\rangle_{ij} \right) \left| 1 \right\rangle_k,
\end{align}
which have the ground-state energy \( e_0 = 18 J^\prime \). When a dimer septet lies on a \(J\) bond, the spin states for an arbitrary spin at site \(k\), \(S^z_k = 1\) and 2, is already a ground state of \( h_{\rm iso}^{(2)} (\alpha_r) \). If it is possible to fulfill the condition of having a dimer septet on every \(J\) bond for the Hamiltonian \( H = \sum h_{\rm iso}^{(2)} \), the product of dimer septet states is the ground state, with the eigenenergy \(E_0 = 18 J^\prime N_{\rm t} \), as is already known for spin-\(\frac12\) systems. Here, \( N_{\rm t} \) is the number of isosceles triangles. For \( \frac{J'}{2J} \leqq \frac12 \), the septet dimer state
\begin{align}
  \left| \Psi_{\rm D} \right\rangle = & \prod_{\rm n.n.} \frac{1}{\sqrt{2}}
  \left( \left| 2 1 \right\rangle - \left| 1 2 \right\rangle \right)_{ij},
  \label{eq:septet_dimer_2D}
\end{align}
is at least one of the ground states.

Similar arguments can be applied to an \(S \geqq 2\) model on a lattice for which the ground state is a product of dimer singlet states, such as a one-dimensional Shastry--Sutherland lattice~\cite{PhysRevLett.47.964}, which includes the Majumdar--Ghosh point~\cite{JMathPhys.10.1063}, a maple-leaf lattice~\cite{PhysRevB.105.L180412,PhysRevB.108.L060406}, quasicrystal lattices~\cite{PhysRevB.108.014426}, a three-dimensional orthogonal-dimer lattice~\cite{J.Phys.Condens.Matter.11.L175}, and others.

\section{summary}   
In conclusion, we found fractionally magnetized quantum states, i.e., coexisting ferromagnetic and spin-singlet states, in an \(S=2\) BLBQ model. Especially, we discover a ferromagnetic Haldane state, where the Haldane state and ferromagnetic state coexist.  The applicability of the ferromagnetic quantum states is guaranteed by the rigorous correspondence between eigenstates at \(\alpha_r\), i.e., by the eigensystem embedding. This means that fractionally magnetized quantum states can exist in any dimension and for any lattice in which the ground state of a spin-\(\frac12\) Heisenberg model is anomalous, e.g., an SPT system~\cite{RevModPhys.89.040502,PhysRevB.87.155114,arXiv2307.04788} or a spin-liquid system~\cite{Nature.464.199,Science.332.1173,NatCommun.4.2287}. Similar arguments can be applied to \(S>2\) BLBQ models. Thus, fractionally magnetized quantum states can exist even for models with larger values of \(S\). In addition, quantized states with \(S_{\rm tot}/N = S-\frac12, S-1, S-\frac32, \dots\) can be stable in a certain parameter range in the model, such as the magnetization plateau. In this way, the present result can lead to the discovery of other types of quantum states, which may be important for both fundamental physics and quantum computer science.

In this paper, we restricted ourselves to the special parameter point \(\alpha_r\) to show that a ferromagnetic quantum state can be realized in a somewhat rigorous way. However, such a ferromagnetic quantum state may be a ground state around \(\alpha_r\) and/or in a model with anisotropic terms. The future arguments of the stability of the ferromagnetic quantum state will be important to find the materials where a ferromagnetic quantum state is realized. Moreover, a ferromagnetic quantum state can be realized even in other spin systems. Thus, the discovery of a ferromagnetic quantum state in more simple models will open the experimental realization. On the other hand, ferromagnetic Haldane state and ferromagnetic dimer multiplet show an anomalous magnetization curve. Since the lowest excitation energy for the \(\Delta S_{\rm tot} = 1 \) process is gapped excitation, magnetization remains constant to be \(M=1-\frac{1}{2S} \) up to the critical external magnetic field \(H_{\rm c}\) corresponding to the spin gap energy. However, the ferromagnetic component increases in the same way as that in the corresponding spin-\(\frac12\) Heisenberg model by applying the external field above \(H_{\rm c}\) as shown in Fig.~\ref{fig:M_H}. Such magnetization is a peculiar signal of any ferromagnetic quantum state. Thus, the experimental observation of such a magnetization curve will open another stage to study the ferromagnetic quantum state.

\begin{acknowledgments}
The authors thank Hosho Katsura for stimulating discussions.
This work was supported by Japan Society for the Promotion of Science (JSPS) KAKENHI Grants No. JP22H01171.
\end{acknowledgments}

  \bibliography{BLBQ}
\end{document}